\documentclass{article}%
\usepackage{amsmath}
\usepackage{amsfonts}
\usepackage{amssymb}
\usepackage{graphicx}%
\providecommand{\U}[1]{\protect\rule{.1in}{.1in}}
\newtheorem{theorem}{Theorem}
\newtheorem{acknowledgement}[theorem]{Acknowledgement}

\begin{document}
\begin{titlepage}
\vspace{.3cm} \vspace{1cm}
\begin{center}
\baselineskip=16pt \centerline{\Large\bf  Who Ordered the Anti-de Sitter Tangent Group? } \vspace{2truecm} \centerline{\large\bf Ali H.
Chamseddine$^{1,2}$\ , \ Viatcheslav Mukhanov$^{3,4,5}$\ \ } \vspace{.5truecm}
\emph{\centerline{$^{1}$Physics Department, American University of Beirut, Lebanon}}
\emph{\centerline{$^{2}$I.H.E.S. F-91440 Bures-sur-Yvette, France}}
\emph{\centerline{$^{3}$Theoretical Physics, Ludwig Maxmillians University,Theresienstr. 37, 80333 Munich, Germany }}
\emph{\centerline{$^{4}$LPT de l'Ecole Normale Superieure, Chaire Blaise Pascal, 24 rue Lhomond, 75231 Paris cedex, France}}
\emph{\centerline{$^{4}$MPI for Physics, Foehringer Ring, 6, 80850, Munich, Germany}}
\end{center}
\vspace{2cm}
\begin{center}
{\bf Abstract}
\end{center}
General relativity can be unambiguously formulated with Lorentz, de Sitter and anti-de Sitter tangent groups, which determine
the fermionic representations. We show that besides of the Lorentz group only anti-de Sitter tangent group is consistent
with all physical requirements.
\end{titlepage}

\bigskip The existence of fermions forces us to consider the tangent group
$SO(3,1),$ or equivalently the $SL\left(  2,\mathbb{C}\right)  $ group, in
General Relativity. Promoting the global Lorentz invariance of the Dirac
action to a local one is achieved by introducing the spin-connection as the
gauge field of the tangent group. The vierbein, which is the square root of
the metric, is the soldering form connecting coordinate basis vectors to
orthogonal tangent vectors. Metricity condition is expressed as the vanishing
of the covariant derivative on the vierbein with respect to both the
spin-connection and the symmetric affine connection. For the Lorentz tangent
group the resulting system of equations is enough to determine both the
spin-connection and the symmetric affine connection in terms of the vierbein
and its first derivatives unambiguously. It was established long ago that in
this case the scalar curvature constructed of the metric manifold is
equivalent to the scalar curvature constructed in terms of the
spin-connection. Hence, in the presence of spinors one can formulate
gravitational interactions in terms of the vierbein and (dependent)
spin-connections and we are assured that the corresponding curvature scalar is
equivalent to the metric dependent scalar curvature \cite{utiyama},
\cite{Kibble}, \cite{Wein}. It came as a surprise that the metricity
conditions can also be unambiguously solved for the de Sitter and anti-de
Sitter tangent groups and the curvature invariants with respect to the tangent
group gauge fields are identical to the metric curvature in these cases
\cite{tangent}. To find the coupling to the spinors for these tangent groups
one must first determine the irreducible spinor representations of the groups
$SO(4,1)$ or $SO(3,2)$ and check when it is possible to construct a realistic
model for the quarks and leptons.

Let us define the inverse soldering forms $e_{A}^{\mu}$ $\ $where
$\mu=0,1,2,3$ is a space-time index and $A=a,4$ with $a=0,1,2,3$ is a tangent
space index. The inverse metric $g^{\mu\nu}$ is then given by $g^{\mu\nu
}=e_{A}^{\mu}\eta^{AB}e_{B}^{\nu}$ with $\eta^{AB}=\left(  \eta^{ab},\eta
^{44}=\epsilon\right)  ,$ where $\eta^{ab}=\mathrm{diag}\left(
1,-1,-1,-1\right)  $ is the Minkowski metric, and $\epsilon=-1$ for the de
Sitter tangent group $SO(4,1)$ and $\epsilon=+1$ for anti-de Sitter group
$SO(3,2).$ The metricity condition is then
\begin{equation}
\nabla_{\mu}e_{A}^{\nu}\equiv\partial_{\mu}e_{A}^{\nu}+\omega_{\mu A}^{\quad
B}e_{B}^{\nu}+\Gamma_{\mu\rho}^{\nu}e_{A}^{\rho}=0, \label{metricity}%
\end{equation}
where $\omega_{\mu A}^{\quad B}$ is the gauge connection of the groups
$SO(4,1)$ or $SO(3,2)$ and $\Gamma_{\mu\rho}^{\nu}$ is the symmetric affine
connection of the diffeomorphism group satisfying $\Gamma_{\mu\rho}^{\nu
}=\Gamma_{\rho\mu}^{\nu}.$ The eighty conditions (\ref{metricity}) can be
solved to determine the eighty fields $\omega_{\mu A}^{\quad B}$ and
$\Gamma_{\mu\rho}^{\nu}$ in terms of $e_{A}^{\nu}$ and the first derivatives
$\partial_{\mu}e_{A}^{\nu}.$ In particular the affine connection is given by
the familiar Christoffel symbols
\begin{equation}
\Gamma_{\mu\rho}^{\nu}=\frac{1}{2}g^{\nu\sigma}\left(  \partial_{\mu}%
g_{\sigma\rho}+\partial_{\rho}g_{\mu\sigma}-\partial_{\sigma}g_{\mu\rho
}\right)  ,
\end{equation}
where $g_{\mu\nu}$ is the inverse of $g^{\mu\nu}.$ The curvature of the
spin-connection $\omega_{\mu A}^{\quad B}$ is given by
\begin{equation}
R_{\mu\nu}^{\hspace{0.05in}\hspace{0.05in}AB}\left(  \omega\right)
=\partial_{\mu}\omega_{\nu}^{\,\,\,AB}-\partial_{\nu}\omega_{\mu}%
^{\,\,\,AB}+\omega_{\mu}^{\,\,\,AC}\omega_{\nu C}^{\quad B}-\omega_{\nu
}^{\,\,\,AC}\omega_{\mu C}^{\quad B},
\end{equation}
and is related to the curvature of the affine connection by the relations
\begin{align}
\,R\left(  \omega\right)   &  =e_{A}^{\mu}R_{\mu\nu}^{\hspace{0.05in}%
\hspace{0.05in}AB}\left(  \omega\right)  e_{B}^{\nu}=-R_{\,\,\,\sigma\mu\nu
}^{\nu}\left(  \Gamma\right)  e^{\sigma A}e_{A}^{\mu}\nonumber\\
&  =R_{\,\,\,\sigma\nu\mu}^{\nu}\left(  \Gamma\right)  g^{\sigma\mu}=R\left(
\Gamma\right)  ,
\end{align}
where
\begin{equation}
R_{\,\,\,\sigma\mu\nu}^{\rho}\left(  \Gamma\right)  =\partial_{\mu}\Gamma
_{\nu\sigma}^{\rho}-\partial_{\nu}\Gamma_{\mu\sigma}^{\rho}+\Gamma_{\mu\kappa
}^{\rho}\Gamma_{\nu\sigma}^{\kappa}-\Gamma_{\nu\kappa}^{\rho}\Gamma_{\mu
\sigma}^{\kappa}.\nonumber
\end{equation}
Thus, although the scalar curvature of the spin-connection is a function of
$e_{A}^{\mu},$ it is expressible as function of the metric $g^{\mu\nu}$ only
because of the invariance under the local $SO(4,1)$ or $SO(3,2)$ gauge
transformations. Thus, in the absence of matter couplings the gravity
formulations in terms of the metric tensor, the vierbein $e_{a}^{\mu}$ for the
Lorentz tangent group $SO(3,1)$, the vielbein $e_{A}^{\mu}$ for the tangent
groups $SO(4,1)$ or $SO(3,2)$ are all equivalent. To find the physical
consequences of the choice of tangent group let us consider matter couplings.

It is well known that the bosonic fields do not \textquotedblleft
feel\textquotedblright\ the tangent group. The relevant fundamental couplings
in this case are those of scalars and vectors. In the case of the Lorentz
tangent group $SO(3,1)$ there is one to one correspondence between vectors
with respect to the diffeomorphism transformations and to the local tangent
group gauge transformations. For the other tangent groups $SO(4,1)$ and
$SO(3,2)$ a five dimensional vector $V_{A}$ with respect to the tangent group
projects into a diffeomorphism vector $A_{\mu}$ and a scalar $\phi$%
\begin{equation}
V_{A}=e_{A}^{\mu}A_{\mu}+n_{A}\phi,
\end{equation}
where $n_{A}$ is a unit vector orthogonal to $e_{A}^{\mu}$%
\begin{equation}
n^{A}e_{A}^{\mu}=0,\qquad n_{A}n^{A}=\varepsilon,
\end{equation}
with $\varepsilon=1$ for $SO(3,2)$ and $\varepsilon=-1$ for $SO(4,1)$
\cite{tangent}. Therefore, there is no obvious advantages in formulating the
vector interactions for the tangent group vectors because this is a reducible
representation and the resulting action does not have any extra symmetries.
Moreover, only the tangent group $SO(4,1)$ gives the correct sign for the
kinetic energy of the extra scalar field which inevitably emerges in this
case. Hence it seems that the only relevant symmetry for the bosonic fields is
the diffeomorphism symmetry.

The situation is different for the fermions as those ones are defined as
representations of the tangent group. For the $SO(3,1)$ tangent group the
spinors transform under local gauge transformations as%
\begin{equation}
\delta\psi_{\alpha}=\frac{1}{4}\Lambda_{ab}\left(  \gamma^{ab}\right)
_{\alpha}^{\beta}\psi_{\beta},\qquad a=0,1,2,3,
\end{equation}
where $\Lambda_{ab}=-\Lambda_{ba}$ are the gauge parameters and $\gamma
^{ab}=\frac{1}{2}\left(  \gamma^{a}\gamma^{b}-\gamma^{b}\gamma^{a}\right)  .$
We adopted here the notation where the Dirac matrices satisfy $\left\{
\gamma^{a},\gamma^{b}\right\}  =2\eta^{ab}$ with
\begin{equation}
\eta^{ab}=\mathrm{diag}\left(  1,-1,-1,-1\right)  .
\end{equation}
The transformations preserve either Weyl or Majorana condition on the spinors
but not both ones. If the Weyl condition $\gamma_{5}\psi=\psi$ is imposed,
where $\gamma_{5}=i\gamma^{0}\gamma^{1}\gamma^{2}\gamma^{3}$ satisfies
$\left(  \gamma_{5}\right)  ^{2}=1,$ then
\begin{equation}
\delta\left(  \gamma_{5}\psi\right)  _{\alpha}=\frac{1}{4}\Lambda_{ab}\left(
\gamma^{ab}\right)  _{\alpha}^{\beta}\left(  \gamma_{5}\psi\right)  _{\beta}.
\end{equation}
Alternatively, imposing the Majorana condition $\psi_{\alpha}=C_{\alpha\beta
}\left(  \overline{\psi}^{\beta}\right)  ^{T}$ where $C$ is the charge
conjugation matrix, is preserved by the gauge transformations because it is
possible to find  Dirac matrices, satisfying the symmetry condition $\left(
\gamma^{ab}C\right)  _{\alpha\beta}=\left(  \gamma^{ab}C\right)  _{\beta
\alpha}$ \cite{Scherk}. We note that the Standard Model is formulated in terms
of chiral leptons and quarks, while the minimally supersymmetric standard
model is formulated in terms of Majorana fermions. This is possible because
all physical fermions correspond to Dirac spinors which acquire their mass by
coupling left-handed spinors to the right handed ones, which in turn can also
be decomposed as the complex sum of two Majorana spinors. Although direct
Majorana mass terms are possible, these would break the $SU(3)\times
SU(2)\times U(1)$ gauge symmetries, except for the right-handed neutrino,
which is desirable.

Repeating this analysis for the $SO(4,1)$ tangent group, the gauge
transformations of the spinors are now
\begin{equation}
\delta\psi_{\alpha}=\frac{1}{4}\Lambda_{AB}\left(  \Gamma^{AB}\right)
_{\alpha}^{\beta}\psi_{\beta},\text{\qquad}A=a,4,
\end{equation}
where the Dirac matrices $\Gamma^{A}$ satisfy $\left\{  \Gamma^{A},\Gamma
^{B}\right\}  =2\eta^{AB}$ where
\[
\eta^{AB}=\mathrm{diag}\left(  1,-1,-1,-1,-1\right)  .
\]
A\ convenient representation is to take $\Gamma^{a}=\gamma^{a}$ and
$\Gamma^{4}=i\gamma_{5}$. In this case it is obvious that the Weyl condition
$\gamma_{5}\psi=\psi$ is not preserved by the gauge transformation. In
addition, the Majorana condition $\psi_{\alpha}=C_{\alpha\beta}\left(
\overline{\psi}^{\beta}\right)  ^{T}$ is also not preserved by the gauge
transformations because
\begin{equation}
\left(  i\gamma_{5}\gamma_{a}\psi\right)  =-C\overline{\left(  i\gamma
_{5}\gamma_{a}\psi\right)  }^{T}.
\end{equation}
The minus sign is a consequence of the signature $\eta^{44}=-1$ which forces
$\Gamma^{4}$ to be equal to $i\gamma_{5}.$ This also implies that the Majorana
condition could be imposed for the tangent group $SO(3,2).$ Thus, when the
tangent group is $SO(4,1)$ only complex Dirac spinors are allowed. To
construct realistic models, each of the fermionic representations must be
complex, and for the Standard Model the lepton doublet would have both right
and left-handed components, and similarly for the leptonic singlets. This
creates the problem of mirror fermions, as the physical particle formed by
coupling the Higgs field to the leptonic doublets and singlet, will have a
partner with the same mass but formed from the opposite chiralities. For
example if we denote the Dirac leptonic doublet  by $l=\left(
\begin{array}
[c]{c}%
\nu_{e}\\
e
\end{array}
\right)  $, the Dirac singlet by $\widetilde{e}$ and the Higgs field by $H$
\ the fermionic terms are then given by
\begin{align}
&  i\overline{l}e_{A}^{\mu}\Gamma^{A}\left(  \partial_{\mu}+\frac{1}{4}%
\omega_{\mu}^{BC}\Gamma_{BC}\right)  l+i\overline{\widetilde{e}}e_{A}^{\mu
}\Gamma^{A}\left(  \partial_{\mu}+\frac{1}{4}\omega_{\mu}^{BC}\Gamma
_{BC}\right)  \widetilde{e}\\
&  +f\left(  \overline{l}H\widetilde{e}+\overline{\widetilde{e}}H^{\ast
}l\right)  +m\left(  \overline{l}l+\overline{\widetilde{e}}\,\widetilde{e}%
\right)  .
\end{align}
When $H$ gets an expectation value, a mass term results of the form
\begin{equation}
f\mu\left(  \overline{e}_{L}\widetilde{e}_{R}+\overline{e}_{R}\widetilde{e}%
_{L}+\overline{\widetilde{e}}_{L}e_{R}+\overline{\widetilde{e}}_{R}%
e_{L}\right)  +m\left(  \overline{e}_{L}e_{R}+\overline{e}_{R}e_{L}%
+\overline{\widetilde{e}}_{L}\widetilde{e}_{R}+\overline{\widetilde{e}}%
_{R}\widetilde{e}_{L}\right)  ,
\end{equation}
which shows that we have two massive electrons formed from combinations of
$e_{L}$, $\widetilde{e}_{R}$ and $e_{R}$, $\widetilde{e}_{L}.$ One must then
tune one combination to have a small mass and identified with the electron
while the other combination would be heavy. In the Standard Model because of
quadratic divergencies in the Higgs sector, the fermionic masses must be tuned
to be low, which is the hierarchy problem. In this case fine tuning is needed
between the two fermion masses to keep one combination low, in addition to the
hierarchy problem resulting from quadratic divergencies. In this respect, the
tangent group $SO(4,1)$ seems to be less natural than the $SO(3,1)$ tangent group.

We can easily see that when the tangent group is $SO(3,2)$ we can impose the
Majorana condition. It is known that the algebra $SO(3,2)$ is isomorphic to
the algebra $SP\left(  2,2\right)  $ with generators $M_{\alpha\beta}%
=M_{\beta\alpha}$ satisfying the commutation relations \cite{CW}, \cite{MM},
\cite{chams},
\begin{equation}
\left[  M_{\alpha\beta},M_{\gamma\delta}\right]  =C_{\alpha\gamma}%
M_{\beta\delta}+C_{\beta\gamma}M_{\alpha\delta}+C_{\alpha\delta}M_{\beta
\gamma}+C_{\beta\delta}M_{\alpha\gamma},
\end{equation}
where $C_{\alpha\beta}=-C_{\beta\alpha}$ is the charge conjugation matrix.
Connection between the generators of the two groups is made through the
identification $M_{\alpha\beta}=\frac{1}{4}M_{AB}\left(  \Gamma^{AB}C\right)
_{\alpha\beta},$ where $A=a,4$ and $\Gamma^{a}=i\gamma_{5}\gamma^{a},$
$\Gamma^{4}=\gamma_{5}.$ The gauge transformations then take the simple form
$\delta\psi_{\alpha}=M_{\alpha}^{\beta}\psi_{\beta}$ where $M_{\alpha}^{\beta
}=M_{\alpha\gamma}C^{\gamma\beta}$ and thus preserve the Majorana condition.
The vielbein $e_{A}^{\mu}$ are in the antisymmetric representation of
$SP\left(  2,2\right)  $%
\begin{equation}
\left(  e^{\mu}\right)  _{\alpha\beta}=-\left(  e^{\mu}\right)  _{\beta\alpha
}=\left(  \Gamma^{A}C\right)  _{\alpha\beta}e_{A}^{\mu}%
\end{equation}
The fermionic action can thus be written in terms of the matrices $\left(
e^{\mu}\right)  _{\alpha\beta}$ and $\left(  \omega_{\mu}\right)
_{\alpha\beta}=\frac{1}{4}\omega_{\mu}^{AB}\left(  \Gamma^{AB}C\right)
_{\alpha\beta}$%
\begin{equation}
i\overline{\psi}e^{\mu}\left(  \partial_{\mu}+\omega_{\mu}\right)  \psi
\end{equation}
and the commutator of the covariant derivatives $D_{\mu}=\partial_{\mu}%
+\omega_{\mu}$ gives
\begin{align}
\left[  D_{\mu},D_{\nu}\right]  _{\alpha\beta} &  =R_{\mu\nu\alpha\beta
}\nonumber\\
&  =\frac{1}{4}R_{\mu\nu}^{\quad AB}\left(  \Gamma_{AB}C\right)  _{\alpha
\beta}%
\end{align}
We conclude that there is no obstruction to construct the Standard Model using
the tangent group $SO(3,2)$ and using Majorana spinors in the same way the
minimally supersymmetric standard model is built. The bosons transform under
the diffeomorphism group and do not feel the tangent group. In this respect
there is no advantage in curved spaces of using the Lorentz group instead of
the anti-de Sitter group. This leads to the puzzling question of whether there
is any significance to the ambiguity in having two possibilities for the
tangent group.
\bigskip
\begin{acknowledgement}
The work of AHC is supported in part by the National Science Foundation
\ Phys-0854779, and Phys-1202671. The work of VM is supported by
\textquotedblleft Chaire Internationale de Recherche Blaise Pascal
financ\'{e}e par l'Etat et la R\'{e}gion d'Ile-de-France, g\'{e}r\'{e}e par la
Fondation de l'Ecole Normale Sup\'{e}rieure\textquotedblright, by TRR 33
\textquotedblleft The Dark Universe\textquotedblright\ and the Cluster of
Excellence EXC 153 \textquotedblleft Origin and Structure of the
Universe\textquotedblright. \bigskip
\end{acknowledgement}

\end{document}